\documentstyle[12pt]{article}

\begin{document}
\title{LHC signatures for $Z^{'}$ 
models with continuously distributed mass}
\author{N.V.Krasnikov  
  \\
INR RAN, Moscow 117312}

\maketitle
\begin{abstract}

We discuss phenomenological consequences of renormalizable $Z^{'}$ models with
continuously distributed mass. 
We point out that 
one of  possible LHC signatures for such model is the existence of   
broad  resonance in Drell-Yan reaction 
$pp \rightarrow Z^{'} \rightarrow l^{+}l^{-}$. 

\end{abstract}

\newpage

The aim of this note is the discussion of the LHC signatures for 
 renormalizable models with continuously distributed mass 
proposed in refs.\cite{1,2}. Note that recent notion of an unparticle,   
introduced by Georgi \cite{3,4} can be interpreted as a 
particular case of a field  with continuously distributed mass \cite{1,2,5,6,7}.
Namely, we consider renormalizable models with vector  interactions 
\cite{1,2}. We point out that 
one of  possible LHC signatures for such models is the existence of   
broad  resonance in Drell-Yan reaction 
$pp \rightarrow Z^{'} \rightarrow l^{+}l^{-}$.

Consider 
the Stueckelberg Lagrangian \cite{8}
\begin{equation}
L_0 = \sum_{k=1}^{N}[-\frac{1}{4}F^{\mu\nu,k}F_{\mu\nu,k} 
+\frac{m^2_k}{2}(A_{\mu,k} - \partial_{\mu}\phi_k)^2] \,,
\end{equation}
where $F_{\mu\nu,k} = \partial_{\mu}A_{\nu,k} - \partial_{\nu}A_{\mu,k}$. 
The Lagrangian (1) is invariant under gauge transformations 
\begin{equation}
A_{\mu,k} \rightarrow A_{\mu,k} +\partial_{\mu}\alpha_k \,,
\end{equation}
\begin{equation}
\phi_k \rightarrow \phi_k + \alpha_k \,
\end{equation}
and it describes N free massive vector fields with masses $m_k$.
For the field $B_{\mu} = \sum_{k=1}^N c_kA_{\mu,k}$ the  propagator in 
transverse gauge is 
\begin{equation}
D_{\mu\nu}(p) = (g_{\mu\nu} - \frac{p_{\mu}p_{\nu}}{p^2})
(\sum^N_{k=1} \frac{|c|^2_k}{p^2-m^2_k}) \,. 
\end{equation}
In the limit $N \rightarrow \infty $
\begin{equation}
D_{\mu\nu}(p)  
\rightarrow (g_{\mu\nu} - \frac{p_{\mu}p_{\nu}}{p^2}) D_{int}(p^2)\,, 
\end{equation}
where 
\begin{equation}
D_{int}(p^2)  = \int _0^{\infty}\frac{\rho(t)}{p^2 - t +i\epsilon}dt
\end{equation}
and $\rho(t) = lim_{N\rightarrow \infty }|c^2_k|\delta(t -m^2_k)  \geq 0$. 
One can introduce the interaction of the field $B_{\mu}$ with  
fermion field $\psi$ in standard way, namely 
\begin{equation}
L_{int} = e\bar{\psi}\gamma_{\mu}\psi B^{\mu} \,.
\end{equation}
The Feynman rules for this model coincide with Feynman rules for 
quantum electrodynamics except  the replacement of the photon propagator

\begin{equation}
D^{tr}_ {\mu\nu}(p) = 
(g_{\mu\nu} - \frac{p_{\mu}p_{\nu}}{p^2})\frac{1}{p^2} \rightarrow 
(g_{\mu\nu} - \frac{p_{\mu}p_{\nu}}{p^2})D_{int}(p^2) \,.
\end{equation}
This generalization of quantum electrodynamics 
preserves the renormalizability for finite  
$ \int _0^{\infty}\rho(t)dt$ because the ultraviolet asymptotic of 
$ D_{int}(p^2)$ coincides with  free photon propagator $\frac{1}{p^2}$. 
Note that for $\rho(t) \sim t^{\delta -1}$ we reproduce the case of vector 
unparticle with propagator $\sim\frac{1}{(p^2)^{1-\delta}}$. 
For the propagator $D_{int}(p^2) = \frac{1}{{p^2}} + 
\frac{1}{(p^2  - M^2)} $ we obtain generalization of 
quantum electrodynamics  with  additional massive vector field. 
Consider a model with spectral density 
\begin{equation}
\rho(t) = k(t-t_1)(-t+t_2)
\end{equation}
for $ t_1 \leq t \leq t_2$ and $\rho(t) = 0$ for $t \geq t_2$ ot 
$t \leq t_1$. The coefficient $k$ is determined from the 
normalization condition   $ \int _0^{\infty}\rho(t)dt = 1$
and it is equal to $k = \frac{6}{(t_2 -t_1)^3}$.
For the spectral density (9) the propagator $D_{int}(p^2)$ has 
the form
\begin{equation}
D_{int}(p^2) = k[-\bar{t_1}\bar{t_2} ln \frac{\bar{t_2}}{\bar{t_1}} 
 + \frac{1}{2}(\bar{t_1}^2 - \bar{t_2}^2)  ] \,,
\end{equation}
where $\bar{t}_1  = t_1 - p^2$ and $\bar{t}_2  = t_2 - p^2 $.
Note that in the limit $t_1 \rightarrow t_2 $ the spectral 
density $\rho(t) \rightarrow \delta(t -t_1)$ , 
$D_{int}(p^2) \rightarrow \frac{1}{p^2 -t_1} $
and the model describes  interaction of massive vector field 
with fermions. The propagator (10) of the model does not contain 
any singularity in $p^2$ in comparison with  $ \frac{1}{p^2 -t_1} $ 
propagator which has singularity for $p^2 = t_1$.  The function
$|D_{int}(p^2)|$ has a maximum $\sim \frac{1}{t_2 - t_1}$ in comparison with 
the maximum $\frac{1}{\Gamma m}$ of the propagator 
$|D_{\Gamma}(p^2)| = |\frac{1}{p^2 -m^2 -i\Gamma m}|$. 
Note that the propagator $D_{\Gamma}(p^2)$ 
takes into account the finite decay width of vector boson.   
So vector particle with continuously distributed mass looks 
like standard vector particle with some internal decay width 
which is determined by spectral density $\rho(t)$. 
Consider the second example of the spectral density $\rho(t)$  based on closed analogy with 
the propagator $D_{\Gamma}(p^2)$. Namely, approximately the following equality takes place:
\begin{equation}
\frac{1}{p^2 -m^2 -im\Gamma_{int}} \approx \int_0^{\infty}\frac{\rho(t)dt}{p^2 -t -i\epsilon} \,, 
\end{equation}
where
\begin{equation} 
\rho(t)  =\frac{1}{\pi}\frac{\Gamma_{int} m}{(t -m^2)^2 +\Gamma^2_{int}m^2}.
\end{equation}
For GUT inspired  $Z^{'}$ models \cite{9} the ratio of the total decay width to $Z^{'}$ mass typically 
is $O(1)$ percent. For the $Z_{SSM}$ model \footnote{In the $Z_{SSM}$ model the couplings of $Z^{'}$ boson 
with quarks and leptons coincide with the corresponding couplings of $Z$ boson \cite{9}.} the ratio 
$\frac{\Gamma}{M}$ is the maximal one among GUT inspired models  and it is equal to 
$(\frac{\Gamma}{M})_{SSM} =0.03$. Typical invariant dilepton mass resolutions for Drell-Yan reactions 
\begin{equation}
pp \rightarrow \mu^{+}\mu^{-} + ...  \,,
\end{equation}
\begin{equation}
pp \rightarrow e{+}e{-} + ...
\end{equation}

are 4 percent  (for $M_{Z^{'}} = 1~TeV$, $\mu^{+}\mu^{-}$, CMS detector \cite{10}) and 2 percent 
  (for $M_{Z^{'}} = 1~TeV$, $e^{+}e^{-}$, CMS detector \cite{10}).
\footnote{For the ATLAS detector mass resolutions are similar \cite{11}.} It means that 
for GUT  inspired $Z^{'}$ boson LHC will not be able to measure the decay width of 
$Z^{'}$ boson. For  $Z^{'}$  boson with continuously distributed mass 
 and with internal decay width  (12) $\Gamma_{int}$ bigger than  the $e^{+}e^{-}$ or $\mu^{+}\mu^{-}$ 
invariant mass detector resolutions 
 we can measure the internal decay width $\Gamma_{int}$ and thus distinguish the model 
with continuously distributed mass from GUT inspired $Z^{'}$ models. Note that we can modify the  
$Z^{'}$ model  by the introduction of some additional neutral massive fermion $\nu_{M}$ which interacts 
with $Z^{'}$ boson like $\bar{\nu}_M(g_L\gamma_{\mu}(1 -\gamma_5) + 
g_R\gamma_{\mu}(1 + \gamma_5))Z^{'}_{\mu}$.  For $2 M_{\nu_{M}} 
< M_{Z^{'}}$  the $Z^{'}$ boson has  invisible decays  into two neutral leptons $\nu_{M}$ with 
some decay width $\Gamma_{MM}$ and for big $g_{L}, ~g_{R}$ the decay 
channel $Z^{'} \rightarrow \nu_{M}\bar{\nu}_{M}$  dominates.
This model imitates  the effects related with nonzero internal decay width of
$Z^{'}$ boson with continuously distributed mass. There are two evident observable effects related with 
nonzero $\Gamma_{int}$ for LHC phenomenology. For the Drell-Yan reaction
\begin{equation}
pp \rightarrow Z^{'} \rightarrow l^{+}l^{-}
\end{equation}
the cross section is
\begin{equation}
\sigma(pp \rightarrow  Z^{'} \rightarrow l^{+}l^{-}) = \sigma(pp \rightarrow Z^{'})\cdot 
Br(Z^{'} \rightarrow l^{+}l^{-}) \,.
\end{equation}
For the case when $Z^{'}$ boson has  additional  ''internal'' decay width $\Gamma_{Z^{'}, int}$
we have additional dilution factor in branching due to nonzero    $\Gamma_{Z^{'}, int}$       , namely: 
\begin{equation}
Br(Z^{'} \rightarrow l^{+}l^{-}) \rightarrow Br(Z^{'} \rightarrow l^{+}l^{-})\cdot 
\frac{  \Gamma_{Z^{'}}}{\Gamma_{Z^{'}} + \Gamma_{Z^{'}, int}}  \,. 
\end{equation}
Another additional factor that can complicate LHC discovery  is that for large  
$\Gamma_{Z^{'}, int}$ the $Z^{'}$ boson becomes rather broad that increases the 
averaging interval and  leads to the increase of Drell-Yan background. Really, for GUT inspired 
$Z^{'}$ boson the ratio $\frac{\Gamma_{Z^{'}}}{M_{Z^{'}}}$ is rather small typically less than 0.03.
For instance, for the SSM model we have $\frac{\Gamma_{Z^{'}}}{M_{Z^{'}}} = 0.03$. For LHC for both 
CMS and ATLAS detectors dimuon invariant mass resolution for $M_{inv}(\mu^{+}\mu^{-}) \geq 1~TeV$ 
is bigger or equal  to  3 percent that means in particular that LHC will not be able to measure 
internal decay widths for  GUT inspired    $Z^{'}$ bosons. For relatively big  $\Gamma_{Z^{'}, int}$, 
say for $\frac{\Gamma_{Z^{'}, int}}{M_{Z^{'}}} = 0.1 $ and $M_{Z^{'}} = 1.5~TeV$ LHC will be able to 
test internal structure of the $Z^{'}$ resonance. The LHC discovery of broad vector resonance will be 
an evidence (not proof of course) in favor of internal structure of $Z^{'}$ resonance. 
As an example consider Drell-Yan production of the SSM $Z^{'}$ boson with $M_{Z^{'}}  = 1~TeV $.
The cross section production for such boson is $\sigma(pp \rightarrow Z^{'} \rightarrow 
\mu^{+}\mu^{-}) \approx 86~fb$ \cite{10}. For the $Z^{'}$ boson with internal decay 
width $\frac{\Gamma_{Z^{'}, int}}{M_{Z^{'}}} = 009$ the suppression factor is
$\frac{  \Gamma_{Z^{'}}}{\Gamma_{Z^{'}} + \Gamma_{Z^{'}, int}} = 0.25$ 
and  $\sigma(pp \rightarrow Z^{'} \rightarrow 
\mu^{+}\mu^{-}) \approx 21.5~fb$. The Drell-Yan cross section which is the main background 
for the $Z^{'}$ production is estimated to be \cite{10} 
\begin{equation}
\sigma_{DY}(pp  \rightarrow \mu^{+}\mu^{-}|m_{inv}(\mu^{+}\mu^{-}) \geq 1~TeV) = 6.6~fb
\end{equation}
For the integral luminosity $L_t = 10~fb^{-1}$ the number of signal and background events in
the dimuon mass interval $ m_{inv}(\mu^{+}\mu^{-}) \geq 1~TeV $ are 
$N_S = 215$, $N_B = 66$ and the significance \cite{12} $S_c = 2(\sqrt{N_S +N_B} - 
\sqrt{N_B}) = 17.2$.\footnote{For integral luminosity $L_t = 1~fb^{-1}$ 
$S_c = 5.4$} It means that such resonance if it exists will be discovered at LHC at low 
luminosity stage. Moreover the dimuon resolution for $m_{inv}(\mu^{+}\mu^{-}) = 1.5~TeV$ is 
around 4 percent so we can measure the  decay width of $Z^{'}$ boson. Note that the use of 
the Drell-Yan reaction 
$pp \rightarrow Z^{'} \rightarrow e^{+}e^{-}$ with 
electron-positron pair in final state could be even more promising since the cross sections 
and the branchings are the same as in dimuon case but electron-positron invariant mass 
resolution $m_{inv}(e^{+}e^{-})$ is better than in the dimuon case  \cite{10}. For instance, for 
$m_{inv}(e^{+}e^{-}) = 1.5~TeV$ the electron-positron mass resolution is estimated to be 
around 2.5 percent \cite{10} that will allow to measure the $Z^{'}$ decay width with better accuracy.
   
Note that Current TEVATRON 
experimental bound on $M$ is $M \geq 850~GeV$ for SSM $Z^{'}$ boson \cite{13}. For the 
model with large TEVATRON $\Gamma_{inv}$ bound is much weaker due to dillution factor and broadness of 
the resonance structure.


To conclude in this note we discussed LHC signatures for $Z^{'}$ models with 
continuously distributed mass. One of the possible effects due to nonzero internal decay width of $Z^{'}$ is 
the existence of rather broad resonance structure in Drell-Yan reaction 
$pp \rightarrow Z^{'} \rightarrow l^{+}l^{-}$. 

I am indebted to S.Shmatov for consultations.
This work was supported by the Grant RFBR 07-02-00256,

\end{document}